\documentclass[graybox]{svmult}

\usepackage{mathptmx}       
\usepackage{helvet}         
\usepackage{courier}        
\usepackage{type1cm}        
\usepackage{makeidx}         
\usepackage{graphicx}        
\usepackage{multicol}        
\usepackage[bottom]{footmisc}
\usepackage{bm}
\makeindex             

\begin{document}

\title*{A physical derivation of the Kerr--Newman\\ black hole solution}
\author{Reinhard Meinel}
\institute{Reinhard Meinel \at Theoretisch-Physikalisches Institut, University
of Jena, Max-Wien-Platz 1, 07743 Jena, Germany,
\email{meinel@tpi.uni-jena.de}}
%
%
\maketitle

\abstract{According to the no-hair theorem, the Kerr--Newman black hole
solution represents the most general asymptotically flat, stationary (electro-)
vacuum black hole solution in general relativity. The procedure described here
shows how this solution can indeed be constructed as the unique solution to the
corresponding boundary value problem of the axially symmetric Einstein--Maxwell
equations in a straightforward manner.}

\section{Introduction: From Schwarzschild to Kerr--Newman}
The Schwarzschild solution, depending on a single parameter (mass $M$),
represents the \textit{general} spherically symmetric vacuum solution to the
Einstein equations. Similarly, the Reissner--Nordstr\"om solution, depending
on two parameters ($M$ and electric charge $Q$), is the \textit{general}
spherically symmetric (electro-) vacuum solution to the Einstein--Maxwell
equations. In contrast, the Kerr--Newman solution,
depending on three parameters ($M$, $Q$ and angular momentum $J$), is only a
\textit{particular} stationary and axially symmetric electro-vacuum solution
to the Einstein--Maxwell equations. However, one can show under quite general
conditions that the Kerr--Newman solution represents the most general
asymptotically flat, stationary electro-vacuum \textit{black hole} solution 
(``no-hair theorem'').
Important contributions to the subject of black hole uniqueness were made by
Israel, Carter, Hawking, Robinson and Mazur (1967--1982), for details see the
recent review~\cite{cch}.

Assuming stationarity and axial symmetry, it is indeed possible to
derive the Kerr--Newman black hole solution in straightforward manner, by
solving the corresponding boundary value problem of the Einstein--Maxwell
equations~\cite{m12}. In the following sections, an outline of this work will
be given. The method is a generalization of the technique developed for
solving a boundary value problem of the vacuum Einstein equations leading
to the global solution describing a uniformly rotating disc of dust
in terms of ultraelliptic functions~\cite{nm95, nm03}, see also~\cite{rfe}.
It is based on the ``integrability'' of
the stationary and axisymmetric vacuum Einstein and electro-vacuum
Einstein--Maxwell equations via the ``inverse scattering method'',
see~\cite{bv01}. In the pure vacuum case, the method was also used to
derive the Kerr black hole solution~\cite{n00, nm03, rfe}.
\section{Einstein--Maxwell equations and related Linear Problem}
The stationary and axisymmetric, electro-vacuum Einstein--Maxwell equations are
equivalent to the Ernst equations~\cite{ernst68}
\begin{equation}
\label{ernst}
f\,\Delta {\mathcal E}=(\nabla {\mathcal E} +
2\bar{\Phi}\nabla \Phi)\cdot\nabla {\mathcal E}\, , \quad
f\,\Delta \Phi=(\nabla {\mathcal E} +2\bar{\Phi}\nabla \Phi)\cdot\nabla \Phi
\end{equation}
\begin{equation}
\label{f}
\mbox{with} \quad f\equiv\Re\, {\mathcal E} + |\Phi|^2\, , \quad
\Delta=\frac{\partial^2}{\partial\rho^2}+
\frac{1}{\rho}\frac{\partial}{\partial\rho}
+\frac{\partial^2}{\partial\zeta^2} \, , \quad \nabla
=(\frac{\partial}{\partial\rho},\frac{\partial}{\partial\zeta}).
\end{equation}
The line element reads
\begin{equation}
\D s^2=f^{-1}[\,h(\D\rho^2+\D\zeta^2)+\rho^2\D\phi^2]
-f(\D t+A\,\D\phi)^2,
\end{equation}
where the coordinates $t$ and $\phi$ are adapted to the Killing vectors
corresponding to stationarity and axial symmetry:
\begin{equation}
\boldsymbol\xi=\frac{\partial}{\partial t}, \quad
\boldsymbol\eta=\frac{\partial}{\partial
\phi}.
\end{equation}
We assume an asymptotic behaviour as \, $r\to \infty$ \, ($\rho=r\sin\theta$,
$\zeta=r\cos\theta$) given by
\begin{equation}
\label{as}
\Re\,{\mathcal E} = 1 - \frac{2M}{r} + {\mathcal O}(r^{-2})\, ,
\quad \Im\, {\mathcal E} = -\frac{2J\cos\theta}{r^2} + {\mathcal O}(r^{-3})\, ,
\quad \Phi = \frac{Q}{r} + {\mathcal O}(r^{-2}) 
\end{equation}
corresponding to asymptotic flatness and the absence of a magnetic monopole
term ($Q$ real).
The metric functions $h$ and $A$ can be calculated from the complex Ernst
potentials
$\mathcal E(\rho,\zeta)$ and $\Phi(\rho,\zeta)$ according to
\begin{equation}
\label{h}
(\ln h)_{,z}=\frac{\rho}{f^2}({\mathcal E}_{,z}+2\bar\Phi\Phi_{,z})
(\bar{\mathcal E}_{,z}+2\Phi{\bar \Phi}_{,z}) -
\frac{4\rho}{f}\Phi_{,z}{\bar\Phi}_{,z}\, ,
\end{equation}
\begin{equation}
\label{A}
A_{,z}=\frac{\I\rho}{f^2}[(\Im\, {\mathcal E})_{,z}-
\I{\bar\Phi}\Phi_{,z}+\I\Phi{\bar \Phi}_{,z}] \qquad
\mbox{($r\to\infty$: \, $h\to 1$, $A\to 0$)}.
\end{equation}
Here complex variables
\begin{equation}
z=\rho+\I\zeta, \quad \bar z=\rho-\I\zeta
\end{equation}
have been used instead of $\rho$ and $\zeta$. Note that $f$ has already been
given in (\ref{f}). The electromagnetic field tensor
\begin{equation}
F_{ik}=A_{k,i}-A_{i,k}\, , \quad A_i\,\D x^i=A_{\phi}\D\phi+A_t\D t
\end{equation}
can also be obtained from the Ernst potentials:
\begin{equation}
\label{Ai}
A_t=-\Re\, \Phi, \qquad
A_{\phi,z}=A\, A_{t,z}-\frac{\I\rho}{f}(\Im\, \Phi)_{,z}
\qquad \mbox{($r\to\infty$: \, $A_{\phi}\to 0$)}.
\end{equation}

The Ernst equations (\ref{ernst}) can be formulated as the integrability
condition of a related Linear Problem (LP). We use the LP of~\cite{nk83}
in a slightly modified form, which is advantageous in the presence of
ergospheres:
\begin{equation}
\label{LP1}
{\vec Y}_{,z}=\left[\left(\begin{array}{ccc}
b_1 & 0 & c_1 \\
0 & a_1 & 0 \\
d_1 & 0 & 0
\end{array}
\right)+\lambda\left(\begin{array}{ccc}
0 & b_1 & 0 \\
a_1 & 0 & -c_1 \\
0 & d_1 & 0
\end{array}
\right)\right]{\vec Y},
\end{equation}
\begin{equation}
\label{LP2}
{\vec Y}_{,\bar{z}}=\left[\left(\begin{array}{ccc}
b_2 & 0 & c_2 \\
0 & a_2 & 0 \\
d_2 & 0 & 0
\end{array}
\right)+\frac{1}{\lambda}\left(\begin{array}{ccc}
0 & b_2 & 0 \\
a_2 & 0 & -c_2 \\
0 & d_2 & 0
\end{array}
\right)\right]{\vec Y}
\end{equation}

\noindent with
\begin{equation}
\label{lambda}
\lambda=\sqrt{\frac{K-\I{\bar z}}{K+\I z}}\, ,
\end{equation}
\begin{equation}
a_1=\bar{b}_2=\frac{{\mathcal E}_{,z}+2\bar{\Phi}\Phi_{,z}}{2f}\, , \quad
a_2=\bar{b}_1=\frac{{\mathcal E}_{,\bar z}+2\bar{\Phi}\Phi_{,\bar z}}{2f}\, ,
\end{equation}
\begin{equation}
c_1=f\bar{d}_2=\Phi_{,z}\, , \quad c_2=f\bar{d}_1=\Phi_{,\bar z}\, .
\end{equation}
The integrability condition
\begin{equation}
{\vec Y}_{,z\bar z}={\vec Y}_{,\bar z z}
\end{equation}
is equivalent to the Ernst equations. The following points are relevant for
the application of soliton theoretic solution methods:
\begin{itemize}
\item{The $3\times 3$ matrix ${\vec Y}$ depends not only on the coordinates
$\rho$ and $\zeta$ (or $z$ and $\bar z$), but also on the additional complex
``spectral parameter'' $K$.}
\item{Since $\bar K$ does not appear, we can assume
without loss of generality that the elements of
${\vec Y}$ are holomorphic functions of $K$ defined on the two-sheeted Riemann
surface associated with (\ref{lambda}), except from the locations of
possible singularities.}
\item{Each column of ${\vec Y}$ is itself a solution to the LP.
We assume that these three solutions are linearly independent.}
\item{For a given solution $\mathcal E$, $\Phi$ to the Einstein--Maxwell
equations, the solution to the LP can be fixed (normalized) by
prescribing ${\vec Y}$ at some point $\rho_0$, $\zeta_0$ of the $\rho$-$\zeta$
plane as a (matrix) function of $K$ in one of the two sheets of the
Riemann surface.}
\item{${\vec Y}$ can be discussed in general as a unique function of $\rho$,
$\zeta$ and $\lambda$.}
\end{itemize}
Three interesting relations result directly from the structure of the
LP (\ref{LP1}, \ref{LP2}):
\begin{equation}
\label{L1}
\hspace*{-1cm} 
[f(\rho,\zeta)]^{-1}{\det}\,{\vec Y}(\rho,\zeta,\lambda)=C_0(K),
\end{equation}
\begin{equation}
\label{L2}
\hspace*{-1cm} {\vec Y}(\rho,\zeta,-\lambda)=\left(\begin{array}{crc}
1 & 0 & 0 \\
0 & -1 & 0 \\
0 & 0 & 1
\end{array}
\right){\vec Y}(\rho,\zeta,\lambda){\vec C}_1(K),
\end{equation}
\begin{equation}
\label{L3}
\hspace*{-1cm} \left[{\vec Y}(\rho,\zeta,1/{\bar\lambda})\right]^{\dagger}
\left(\begin{array}{ccr}
[f(\rho,\zeta)]^{-1} & 0 & 0 \\
0 & -[f(\rho,\zeta)]^{-1} & 0 \\
0 & 0 & -1
\end{array}
\right){\vec Y}(\rho,\zeta,\lambda) = {\vec C}_2(K),
\end{equation}
where $C_0(K)$ as well as the matrices ${\vec C}_1(K)$ and
${\vec C}_2(K)$ do not depend on $\rho$ and $\zeta$.
\section{Solving the black hole boundary value problem}
After formulating the black hole boundary value problem, we will use the LP
to find its solution. The most important part comprises deriving
the Ernst potentials on the axis of symmetry~\cite{m12}. It is well known that
these ``axis data'' uniquely determine the solution everywhere, see~\cite{he80,
s91}. A straightforward method for obtaining the full solution from
the axis data is based on the analytical properties of ${\vec Y}$ as a function
of $\lambda$ \cite{mr12}.
\subsection{Boundary conditions}
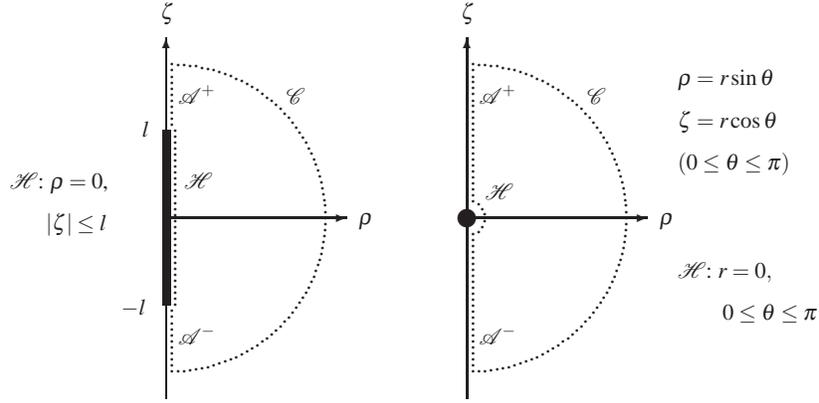
\begin{figure} [t]
\unitlength0.8cm
\begin{picture}(8,7)
\linethickness{1mm}
\put(3.5,1.55){\line(0,1){2.9}}
\thinlines
\put(3.5,3){\vector(1,0){3}}
\put(3.5,0){\vector(0,1){6}}
\put(3.4,6.3){$\zeta$}
\put(6.7,2.9){$\rho$}
\multiput(3.6,0.45)(0,0.1){11}{\circle*{0.02}}
\multiput(3.6,5.55)(0,-0.1){11}{\circle*{0.02}}
\multiput(3.65,1.55)(0,0.1){30}{\circle*{0.02}}
\put(3.70011,5.54803){\circle*{0.02}}
\put(3.80007,5.54214){\circle*{0.02}}
\put(3.89972,5.53232){\circle*{0.02}}
\put(3.99891,5.51861){\circle*{0.02}}
\put(4.09748,5.501){\circle*{0.02}}
\put(4.19529,5.47954){\circle*{0.02}}
\put(4.29217,5.45426){\circle*{0.02}}
\put(4.38799,5.42519){\circle*{0.02}}
\put(4.4826,5.39239){\circle*{0.02}}
\put(4.57584,5.35589){\circle*{0.02}}
\put(4.66758,5.31577){\circle*{0.02}}
\put(4.75768,5.27207){\circle*{0.02}}
\put(4.84598,5.22486){\circle*{0.02}}
\put(4.93237,5.17423){\circle*{0.02}}
\put(5.0167,5.12025){\circle*{0.02}}
\put(5.09885,5.06299){\circle*{0.02}}
\put(5.17869,5.00256){\circle*{0.02}}
\put(5.25609,4.93904){\circle*{0.02}}
\put(5.33094,4.87252){\circle*{0.02}}
\put(5.40312,4.80312){\circle*{0.02}}
\put(5.47252,4.73094){\circle*{0.02}}
\put(5.53904,4.65609){\circle*{0.02}}
\put(5.60256,4.57869){\circle*{0.02}}
\put(5.66299,4.49885){\circle*{0.02}}
\put(5.72025,4.4167){\circle*{0.02}}
\put(5.77423,4.33237){\circle*{0.02}}
\put(5.82486,4.24598){\circle*{0.02}}
\put(5.87207,4.15768){\circle*{0.02}}
\put(5.91577,4.06758){\circle*{0.02}}
\put(5.95589,3.97584){\circle*{0.02}}
\put(5.99239,3.8826){\circle*{0.02}}
\put(6.02519,3.78799){\circle*{0.02}}
\put(6.05426,3.69217){\circle*{0.02}}
\put(6.07954,3.59529){\circle*{0.02}}
\put(6.101,3.49748){\circle*{0.02}}
\put(6.11861,3.39891){\circle*{0.02}}
\put(6.13232,3.29972){\circle*{0.02}}
\put(6.14214,3.20007){\circle*{0.02}}
\put(6.14803,3.10011){\circle*{0.02}}
\put(6.15,3.){\circle*{0.02}}
\put(6.14803,2.89989){\circle*{0.02}}
\put(6.14214,2.79993){\circle*{0.02}}
\put(6.13232,2.70028){\circle*{0.02}}
\put(6.11861,2.60109){\circle*{0.02}}
\put(6.101,2.50252){\circle*{0.02}}
\put(6.07954,2.40471){\circle*{0.02}}
\put(6.05426,2.30783){\circle*{0.02}}
\put(6.02519,2.21201){\circle*{0.02}}
\put(5.99239,2.1174){\circle*{0.02}}
\put(5.95589,2.02416){\circle*{0.02}}
\put(5.91577,1.93242){\circle*{0.02}}
\put(5.87207,1.84232){\circle*{0.02}}
\put(5.82486,1.75402){\circle*{0.02}}
\put(5.77423,1.66763){\circle*{0.02}}
\put(5.72025,1.5833){\circle*{0.02}}
\put(5.66299,1.50115){\circle*{0.02}}
\put(5.60256,1.42131){\circle*{0.02}}
\put(5.53904,1.34391){\circle*{0.02}}
\put(5.47252,1.26906){\circle*{0.02}}
\put(5.40312,1.19688){\circle*{0.02}}
\put(5.33094,1.12748){\circle*{0.02}}
\put(5.25609,1.06096){\circle*{0.02}}
\put(5.17869,0.997442){\circle*{0.02}}
\put(5.09885,0.937007){\circle*{0.02}}
\put(5.0167,0.879752){\circle*{0.02}}
\put(4.93237,0.825768){\circle*{0.02}}
\put(4.84598,0.775135){\circle*{0.02}}
\put(4.75768,0.727933){\circle*{0.02}}
\put(4.66758,0.684235){\circle*{0.02}}
\put(4.57584,0.644107){\circle*{0.02}}
\put(4.4826,0.607612){\circle*{0.02}}
\put(4.38799,0.574806){\circle*{0.02}}
\put(4.29217,0.545739){\circle*{0.02}}
\put(4.19529,0.520457){\circle*{0.02}}
\put(4.09748,0.498998){\circle*{0.02}}
\put(3.99891,0.481395){\circle*{0.02}}
\put(3.89972,0.467675){\circle*{0.02}}
\put(3.80007,0.457861){\circle*{0.02}}
\put(3.70011,0.451966){\circle*{0.02}}
\put(3.7,4.9){$\mathcal{A^+}$}
\put(3.7,0.9){$\mathcal{A^-}$}
\put(5.5,4.9){$\mathcal{C}$}
\put(3.1,4.35){$l$}
\put(2.75,1.4){$-l$}
\put(3.8,3.5){${\mathcal{H}}$}
\put(0.9,3.5){${\mathcal{H}}$: $\rho=0$,}
\put(1.5,2.8){$|\zeta|\le l$}
\put(8.5,3){\vector(1,0){3}}
\put(8.5,0){\vector(0,1){6}}
\put(8.4,6.3){$\zeta$}
\put(11.7,2.9){$\rho$}
\put(8.5,3){\circle*{0.3}}
\put(8.69284,3.22981){\circle*{0.02}}
\put(8.75981,3.15){\circle*{0.02}}
\put(8.79544,3.05209){\circle*{0.02}}
\put(8.79544,2.94791){\circle*{0.02}}
\put(8.75981,2.85){\circle*{0.02}}
\put(8.69284,2.77019){\circle*{0.02}}
\multiput(8.6,0.45)(0,0.099){24}{\circle*{0.02}}
\multiput(8.6,5.55)(0,-0.099){24}{\circle*{0.02}}
\put(8.70011,5.54803){\circle*{0.02}}
\put(8.80007,5.54214){\circle*{0.02}}
\put(8.89972,5.53232){\circle*{0.02}}
\put(8.99891,5.51861){\circle*{0.02}}
\put(9.09748,5.501){\circle*{0.02}}
\put(9.19529,5.47954){\circle*{0.02}}
\put(9.29217,5.45426){\circle*{0.02}}
\put(9.38799,5.42519){\circle*{0.02}}
\put(9.4826,5.39239){\circle*{0.02}}
\put(9.57584,5.35589){\circle*{0.02}}
\put(9.66758,5.31577){\circle*{0.02}}
\put(9.75768,5.27207){\circle*{0.02}}
\put(9.84598,5.22486){\circle*{0.02}}
\put(9.93237,5.17423){\circle*{0.02}}
\put(10.0167,5.12025){\circle*{0.02}}
\put(10.09885,5.06299){\circle*{0.02}}
\put(10.17869,5.00256){\circle*{0.02}}
\put(10.25609,4.93904){\circle*{0.02}}
\put(10.33094,4.87252){\circle*{0.02}}
\put(10.40312,4.80312){\circle*{0.02}}
\put(10.47252,4.73094){\circle*{0.02}}
\put(10.53904,4.65609){\circle*{0.02}}
\put(10.60256,4.57869){\circle*{0.02}}
\put(10.66299,4.49885){\circle*{0.02}}
\put(10.72025,4.4167){\circle*{0.02}}
\put(10.77423,4.33237){\circle*{0.02}}
\put(10.82486,4.24598){\circle*{0.02}}
\put(10.87207,4.15768){\circle*{0.02}}
\put(10.91577,4.06758){\circle*{0.02}}
\put(10.95589,3.97584){\circle*{0.02}}
\put(10.99239,3.8826){\circle*{0.02}}
\put(11.02519,3.78799){\circle*{0.02}}
\put(11.05426,3.69217){\circle*{0.02}}
\put(11.07954,3.59529){\circle*{0.02}}
\put(11.101,3.49748){\circle*{0.02}}
\put(11.11861,3.39891){\circle*{0.02}}
\put(11.13232,3.29972){\circle*{0.02}}
\put(11.14214,3.20007){\circle*{0.02}}
\put(11.14803,3.10011){\circle*{0.02}}
\put(11.15,3.){\circle*{0.02}}
\put(11.14803,2.89989){\circle*{0.02}}
\put(11.14214,2.79993){\circle*{0.02}}
\put(11.13232,2.70028){\circle*{0.02}}
\put(11.11861,2.60109){\circle*{0.02}}
\put(11.101,2.50252){\circle*{0.02}}
\put(11.07954,2.40471){\circle*{0.02}}
\put(11.05426,2.30783){\circle*{0.02}}
\put(11.02519,2.21201){\circle*{0.02}}
\put(10.99239,2.1174){\circle*{0.02}}
\put(10.95589,2.02416){\circle*{0.02}}
\put(10.91577,1.93242){\circle*{0.02}}
\put(10.87207,1.84232){\circle*{0.02}}
\put(10.82486,1.75402){\circle*{0.02}}
\put(10.77423,1.66763){\circle*{0.02}}
\put(10.72025,1.5833){\circle*{0.02}}
\put(10.66299,1.50115){\circle*{0.02}}
\put(10.60256,1.42131){\circle*{0.02}}
\put(10.53904,1.34391){\circle*{0.02}}
\put(10.47252,1.26906){\circle*{0.02}}
\put(10.40312,1.19688){\circle*{0.02}}
\put(10.33094,1.12748){\circle*{0.02}}
\put(10.25609,1.06096){\circle*{0.02}}
\put(10.17869,0.997442){\circle*{0.02}}
\put(10.09885,0.937007){\circle*{0.02}}
\put(10.0167,0.879752){\circle*{0.02}}
\put(9.93237,0.825768){\circle*{0.02}}
\put(9.84598,0.775135){\circle*{0.02}}
\put(9.75768,0.727933){\circle*{0.02}}
\put(9.66758,0.684235){\circle*{0.02}}
\put(9.57584,0.644107){\circle*{0.02}}
\put(9.4826,0.607612){\circle*{0.02}}
\put(9.38799,0.574806){\circle*{0.02}}
\put(9.29217,0.545739){\circle*{0.02}}
\put(9.19529,0.520457){\circle*{0.02}}
\put(9.09748,0.498998){\circle*{0.02}}
\put(8.99891,0.481395){\circle*{0.02}}
\put(8.89972,0.467675){\circle*{0.02}}
\put(8.80007,0.457861){\circle*{0.02}}
\put(8.70011,0.451966){\circle*{0.02}}
\put(8.8,3.3){$\mathcal{H}$}
\put(8.7,4.9){$\mathcal{A^+}$}
\put(8.7,0.9){$\mathcal{A^-}$}
\put(10.5,4.9){$\mathcal{C}$}
\put(12,5.2){$\rho=r\sin\theta$}
\put(12,4.5){$\zeta=r\cos\theta$}
\put(12,3.8){$(0\le\theta\le\pi)$}
\put(12,2.0){$\mathcal{H}$: $r=0,$}
\put(12.75,1.3) {$0\le\theta\le\pi$}
\end{picture}
\caption{In Weyl ccordinates, the horizon is either a finite interval
or a point on the $\zeta$-axis (adapted from~\cite{m12})}
\label{fig1}
\end{figure}

The event horizon $\mathcal H$ of a stationary and axisymmetric black hole is
characterized by the conditions
\begin{equation}
\label{bc}
\mathcal H: \quad \chi^i\chi_i=0\, ,
\quad \chi^i\eta_i=0\, ,
\end{equation}
where $\chi^i\equiv \xi^i+\Omega\eta^i$ and the constant $\Omega$ is the
``angular velocity of the horizon'' \cite{he73, c73}. Because of
\begin{equation}
\rho^2=(\xi^i\eta_i)^2-\xi^i\xi_i\eta^k\eta_k=(\chi^i\eta_i)^2-
\chi^i\chi_i\eta^k\eta_k
\end{equation}
the horizon must be located on the $\zeta$-axis of our Weyl coordinate system:
\begin{equation}
\mathcal H: \quad \rho=0.
\end{equation}
This results in two possibilities for a connected horizon\footnote{A connected
horizon means a single black hole. We are not interested here in the problem
of multi-black-hole equilibrium states.}:
(i) a finite interval on the $\zeta$-axis and
(ii) a point on the $\zeta$-axis,
see Fig.~\ref{fig1}. Note that the two parts of the symmetry axis,
$\mathcal{A^+}$ and $\mathcal{A^-}$, where the Killing vector $\boldsymbol\eta$
vanishes, are also characterized by $\rho=0$. The black hole boundary value
problem consists of finding a solution that is regular everywhere outside the
horizon and satisfies (\ref{bc}) and (\ref{as}).
\subsection{Axis data}
At $\rho=0$, the branch points $K=\I{\bar z}$ and $K=-\I z$ of
(\ref{lambda}) merge to $K=\zeta$ and for $K\ne \zeta$ holds $\lambda=\pm 1$.
Consequently, the solution to the LP, for $\lambda=+1$, is of the form
\begin{equation}
\label{Apm}
\mathcal{A^{\pm}}: \quad {\vec Y}_{\pm}=
\left(\begin{array}{crr}
\bar{\mathcal E}+2|\Phi|^2 & 1 & \Phi \\
{\mathcal E} &-1 & -\Phi \\
2\bar \Phi & 0 & 1
\end{array}
\right){\vec C}_{\pm},
\end{equation}
\begin{equation}
\mathcal H: \hspace{5mm} {\vec Y}_{\mathrm h}=
\left(\begin{array}{crr}
\bar{\mathcal E}+2|\Phi|^2 & 1 & \Phi \\
{\mathcal E} &-1 & -\Phi \\
2\bar \Phi & 0 & 1
\end{array}
\right){\vec C}_{\mathrm h}. 
\end{equation}
We fix ${\vec C}_+(K)$ by the normalization condition
\begin{equation}
\label{C+}
\lim_{K\to \zeta}{\vec Y}_+(\zeta,K) =\left(\begin{array}{crc}
1 & 1 & 0 \\
1 & -1 & 0 \\
0 & 0 & 1
\end{array}\right)
\qquad
\Rightarrow
\qquad
{\vec C}_+ = \left(\begin{array}{ccc}
F & 0 & 0 \\
G & 1 & L \\
H & 0 & 1
\end{array}\right)
\end{equation} 
and the functions $F(K)$, $G(K)$, $H(K)$ and $L(K)$, for $K=\zeta$, are
given by the potentials $\mathcal E=\mathcal E_+$, $\Phi=\Phi_+$ on
$\mathcal{A^+}$:
\begin{equation}
F(\zeta)=[f_+(\zeta)]^{-1}, \quad G(\zeta)=
\left[|\Phi_+(\zeta)|^2+\I \Im\,\mathcal E_+(\zeta)\right][f_+(\zeta)]^{-1},
\end{equation}
\begin{equation}
H(\zeta)=-2\bar\Phi_+(\zeta)[f_+(\zeta)]^{-1}, \quad L(\zeta)=-\Phi_+(\zeta)
\end{equation}
and, vice versa,
\begin{equation}
\label{axis}
\mathcal E_+(\zeta)=\frac{1-\bar G(\zeta)}{F(\zeta)}\, ,
\quad \Phi_+(\zeta)=-\frac{\bar H(\zeta)}{2F(\zeta)}.
\end{equation}
We can calculate $C_0(K)$, ${\vec C}_1(K)$ and
${\vec C}_2(K)$ of relations (\ref{L1}--\ref{L3}) for our
normalization:
\begin{equation}
\label{C}
C_0=-2F, \quad 
{\vec C}_1=\left(\begin{array}{ccc}
0 & 1 & 0 \\
1 & 0 & 0 \\
0 & 0 & 1
\end{array}\right), \quad
{\vec C}_2=\left(\begin{array}{ccr}
0 & 2F & 0 \\
2F & 0 & 0 \\
0 & 0 & -1
\end{array}\right).
\end{equation}
On $\mathcal{A^+}$, (\ref{L3}) reads
\begin{equation}
\label{K3}
\left[{\vec C}_+(\bar K)\right]^{\dagger}\left(\begin{array}{ccr}
0 & 2 & 0 \\
2 & 0 & 0 \\
0 & 0 & -1
\end{array}\right){\vec C}_+(K)=\left(\begin{array}{ccr}
0 & 2F & 0 \\
2F & 0 & 0 \\
0 & 0 & -1
\end{array}\right).
\end{equation}

From continuity conditions at the ``poles'' of the horizon ($\rho=0$,
$\zeta=\pm l$ or $r=0$, $\theta=0,\pi$; see Fig.~\ref{fig1}) and using the
boundary conditions, one can calculate ${\vec C}_{\mathrm h}(K)$ and
${\vec C}_-(K)$ in terms of ${\vec C}_+(K)$, for details I refer to~\cite{m12}.
Closing the path of integration via infinity (curve $\mathcal C$:
$\rho=R\sin\theta$, $\zeta=R\cos\theta$ with $0\le\theta\le\pi$, $R\to\infty$),
where ${\vec Y}$ is constant because of the LP and (\ref{as}),
but $\lambda$ changes from $\pm 1$ at $\theta=0$ to
$\mp 1$ at $\theta=\pi$, we obtain with (\ref{L2}) and (\ref{C}) an explicit
expression for ${\vec C}_+(K)$ in terms of the parameters $\Omega$, $l$ (with
$l=0$ for a horizon at $r=0$) and the values of the Ernst potentials at the
poles. Using (\ref{axis}), we can calculate $\mathcal E_+$ and $\Phi_+$. The
number of free real parameters is reduced to four as a consequence of the
constraint (\ref{K3}) and to three if no magnetic monopole is allowed.
The final result is
\begin{equation}
\label{FG}
F(K)=\frac{(K-L_1)(K-L_2)}{(K-K_1)(K-K_2)}\, , \quad 
G(K)=\frac{Q^2-2\I J}{(K-K_1)(K-K_2)}\, ,
\end{equation}
\begin{equation}
\label{HL}
H(K)=-\frac{2Q(K-L_1)}{(K-K_1)(K-K_2)}\, , \quad L(K)=-\frac{Q}{K-L_1}
\end{equation}
\begin{equation}
\mbox{with} \quad L_{1/2}=-M\pm\I\frac{J}{M}\, , \quad
K_{1/2}=\pm\sqrt{M^2-Q^2-\frac{J^2}{M^2}}\,
\end{equation}
and, correspondingly,
\begin{equation}
\mathcal E_+(\zeta)=1-\frac{2M}{\zeta+M-\I J/M} \, , \quad
\Phi_+(\zeta)=\frac{Q}{\zeta+M-\I J/M}
\end{equation}
together with the parameter relations
\begin{equation}
\label{par}
\frac{l^2}{M^2}+\frac{Q^2}{M^2}+\frac{J^2}{M^4}=1 \qquad \mbox{and} \qquad
\Omega M=\frac{J/M^2}{(1+l/M)^2+J^2/M^4}.
\end{equation}
\subsection{Solution everywhere outside the horizon}
Relation (\ref{L2}) together with the expression for ${\vec C}_1(K)$
in (\ref{C}) is equivalent to the following structure of $\vec Y$:
\begin{equation}
{\bf Y}(\rho,\zeta,\lambda)=\left(\begin{array}{crc}
\psi(\rho,\zeta,\lambda) & \psi(\rho,\zeta,-\lambda) &
\alpha(\rho,\zeta,\lambda) \\
\chi(\rho,\zeta,\lambda) & -\chi(\rho,\zeta,-\lambda) &
\beta(\rho,\zeta,\lambda) \\
\varphi(\rho,\zeta,\lambda) & \varphi(\rho,\zeta,-\lambda) &
\gamma(\rho,\zeta,\lambda)
\end{array}
\right),
\end{equation}
where $\alpha(\rho,\zeta,\lambda)=\alpha(\rho,\zeta,-\lambda)$,
$\beta(\rho,\zeta,\lambda)=-\beta(\rho,\zeta,-\lambda)$ and
$\gamma(\rho,\zeta,\lambda)=\gamma(\rho,\zeta,-\lambda)$. The general
solution of the LP for $K\to\infty$ and $\lambda=+1$ reads
\begin{equation}
\label{Y1}
\vec Y(\rho,\zeta,1)=
\left(\begin{array}{crr}
\bar{\mathcal E}+2|\Phi|^2 & 1 & \Phi \\
{\mathcal E} &-1 & -\Phi \\
2\bar \Phi & 0 & 1
\end{array}
\right)\vec C,
\end{equation}
where $\vec C$ is a constant matrix. Eqs.~(\ref{Apm}, \ref{C+}, \ref{FG},
\ref{HL}) imply $\vec C=\vec 1$ and lead to the ansatz
\begin{equation}
\psi=1+k_1\left(\frac{1}{\kappa_1-\lambda}-\frac{1}{\kappa_1+1}\right)
+k_2\left(\frac{1}{\kappa_2-\lambda}-\frac{1}{\kappa_2+1}\right),
\end{equation}
\begin{equation}
\chi=1+l_1\left(\frac{1}{\kappa_1-\lambda}-\frac{1}{\kappa_1+1}\right)
+l_2\left(\frac{1}{\kappa_2-\lambda}-\frac{1}{\kappa_2+1}\right),
\end{equation}
\begin{equation}
\varphi=m_1\left(\frac{1}{\kappa_1-\lambda}-\frac{1}{\kappa_1+1}\right)
+m_2\left(\frac{1}{\kappa_2-\lambda}-\frac{1}{\kappa_2+1}\right),
\end{equation}
\begin{equation}
\alpha=\Phi+\frac{\alpha_0}{K-L_1}\, , \quad 
\beta=-\Phi\,\frac{\lambda(K+\I z)}{K-L_1}\, ,
\quad \gamma=1+\frac{\gamma_0}{K-L_1}\, ,
\end{equation}
where
\begin{equation}
\kappa_{\mu}=
\sqrt{\frac{K_{\mu}-\I {\bar z}}{K_{\mu}+\I z}}
\quad (\mathcal{A^+}:\quad \kappa_{\mu}=+1)\, .
\end{equation}

According to the LP, ${\vec Y}_{,z}\vec Y^{-1}$ and
${\vec Y}_{,{\bar z}}\vec Y^{-1}$ must be holomorphic functions of $\lambda$
for all $\lambda\ne 0,\infty$. The regularity at $\lambda=\pm\kappa_{\mu}$
($\mu=1,2$), the poles of the first two columns of $\vec Y$, is automatically 
guarantied, whereas regularity at $\lambda=\pm\lambda_{\mu}$ with
$\lambda_{\mu}=\sqrt{(L_{\mu}-\I{\bar z})/(L_{\mu}+\I z)}$
($\mathcal{A^+}$: $\lambda_{\mu}=+1$), where poles of the
third column \linebreak ($\mu=1$) and zeros of $\det{\vec Y}$ ($\mu=1,2$)
occur, see (\ref{L1}, \ref{C}, \ref{FG}),
is equivalent to a set of linear algebraic equations, which together with
(\ref{Apm}, \ref{C+}, \ref{FG}, \ref{HL}) uniquely determine
the unknowns $k_{\mu}(\rho,\zeta)$, $l_{\mu}(\rho,\zeta)$,
$m_{\mu}(\rho,\zeta)$, $\alpha_0(\rho,\zeta)$, $\gamma_0(\rho,\zeta)$ and
$\Phi(\rho,\zeta)$. With ${\mathcal E}(\rho,\zeta)=\chi(\rho,\zeta,1)$, see
(\ref{Y1}), this leads to the result
\begin{equation}
\mathcal E=1-\frac{2M}{\tilde r-\I(J/M)\cos \tilde\theta} \, , \quad
\Phi=\frac{Q}{\tilde r-\I(J/M)\cos \tilde\theta}
\end{equation}
\begin{equation}
\mbox{with} \quad \rho=\sqrt{\tilde r^2-2M\tilde
r+J^2/M^2+Q^2}\,\sin\tilde\theta \, , \quad
\zeta=(\tilde r-M)\cos\tilde\theta.
\end{equation}

The ``domain of outer communication'' (the region outside the event horizon
$\mathcal H$) is given by
$\tilde r > \tilde r_{\mathrm h}=M+\sqrt{M^2-J^2/M^2-Q^2}$.
The horizon itself is characterized by $\tilde r = \tilde r_{\mathrm h}$,
and the axis of symmetry is located at $\tilde \theta=0$ ($\mathcal{A^+}$)
and $\tilde \theta=\pi$ ($\mathcal{A^-}$). Note that (\ref{par}) implies
$Q^2+J^2/M^2\le M^2$. The equality sign, corresponding to $l=0$, is valid
for the extremal Kerr--Newman black hole.
\subsection{Full metric and electromagnetic field}
Using Eqs.~(\ref{f}, \ref{h}, \ref{A}, \ref{Ai}) we can calculate the full
metric and the electromagnetic four-potential:
\begin{eqnarray}
\D s^2 & = & \frac{\Sigma}{\Delta}\, \D\tilde r^2 +
\Sigma\, \D\tilde\theta^2 +\left(\tilde r^2+a^2+
\frac{(2M\tilde r-Q^2)a^2\sin^2\tilde\theta}{\Sigma}\right)
\sin^2\tilde\theta \, \D\phi^2 \\
     &   & - \; \frac{(2M\tilde r-Q^2)2a\sin^2\tilde\theta}{\Sigma}\,
\D\phi\, \D t -\left(1-\frac{2M\tilde r-Q^2}{\Sigma}\right)\D t^2
\end{eqnarray}
\begin{equation}
\mbox{with} \quad \Delta = \tilde r^2-2M\tilde r+a^2+Q^2, \quad
\Sigma = \tilde r^2+a^2\cos^2\tilde\theta, \quad a\equiv J/M
\end{equation}
and
\begin{equation}
A_i\,\D x^i=\frac{Q\tilde r}{\Sigma}(a\sin^2\tilde\theta\,\D\phi-\D t).
\end{equation}
This is the well-known Kerr--Newman solution in Boyer--Lindquist coordinates
$\tilde r$ and $\tilde\theta$. For $Q=0$ it reduces to the
Kerr solution, $J=0$ gives the Reissner--Nordstr\"om solution and $Q=J=0$ leads
back to the Schwarzschild solution.

\end{document}